\documentclass{appolb}
 \usepackage[dvips]{graphics}
  \usepackage{epsfig,epsf,amsmath,amssymb,bm}
   \graphicspath{{./figs-bakulev/}}

\begin{document}

\title{Resummation in Fractional APT:\\
        How many loops do we need to take into account?
       \thanks{Presented at the International Meeting ``Excited QCD'', February 8--14, 2009,
               Zakopane (Poland)}}

\author{Alexander~P.~Bakulev
 \address{Bogoliubov Lab. of Theoretical Physics, JINR \\
          Dubna 141980, Russia\\
          E-mail: bakulev@theor.jinr.ru}}
\maketitle

\begin{abstract}
\noindent We give a short introduction to the Analytic Perturbation Theory (APT)~\cite{SS}
 and its generalization to Fractional powers --- FAPT~\cite{BMS-APT,BKS05}.
 We describe how to treat heavy-quark thresholds in FAPT and then show how
 to resum perturbative series in both the one-loop APT and FAPT.
 As an application we consider FAPT description of the Higgs boson decay $H^0\to b\bar{b}$.
\end{abstract}

\markboth{\large \sl \underline{Alexander~P.~Bakulev}
          \hspace*{2cm} Excited QCD, 2009}
         {\large \sl \hspace*{1cm} Resummation in FAPT: How many loops to calculate?}

\section{APT and FAPT in QCD}
 \label{sec:APT}
In the standard QCD Perturbation Theory (PT) we know
the Renormalization Group (RG) equation $da_s[L]/dL = -a_s^2-\ldots$
for the effective coupling $\alpha_s(Q^2)=a_s[L]/\beta_f$
with $L=\ln(Q^2/\Lambda^2)$, $\beta_f=b_0(N_f)/(4\pi)=(11-2N_f/3)/(4\pi)$\footnote{%
   We use notations $f(Q^2)$ and $f[L]$ in order to specify the arguments we mean ---
   squared momentum $Q^2$ or its logarithm $L=\ln(Q^2/\Lambda^2)$,
   that is $f[L]=f(\Lambda^2\cdot e^L)$ and $\Lambda^2$ is usually referred to $N_f=3$ region.}.
Then the one-loop solution generates Landau pole singularity,
$a_s[L] = 1/L$.

In the Analytic Perturbation Theory (APT) we have
different effective couplings in Minkowskian
(Radyushkin \cite{Rad82}, and Krasnikov and Pivovarov \cite{KP82})
and Euclidean (Shirkov and Solovtsov \cite{SS}) regions.
In Euclidean domain,
$\displaystyle-q^2=Q^2$, $\displaystyle L=\ln Q^2/\Lambda^2$,
APT generates the following set of images for the effective coupling
and its $n$-th powers,
$\displaystyle\left\{{\mathcal A}_n[L]\right\}_{n\in\mathbb{N}}$,
whereas in Minkowskian domain,
$\displaystyle q^2=s$, $\displaystyle L_s=\ln s/\Lambda^2$,
it generates another set,
$\displaystyle\left\{{\mathfrak A}_n[L_s]\right\}_{n\in\mathbb{N}}$.
APT is based on the RG and causality
that guaranties standard perturbative UV asymptotics
and spectral properties.
Power series $\sum_{m}d_m a_s^m[L]$
transforms into non-power series
$\sum_{m}d_m {\mathcal A}_{m}[L]$ in APT.

By the analytization in APT for an observable $f(Q^2)$
we mean the ``K\"allen--Lehman'' representation
 \begin{eqnarray}
  \label{eq:An.SD}
  \left[f(Q^2)\right]_\text{an}
   = \int_0^{\infty}\!
      \frac{\rho_f(\sigma)}
         {\sigma+Q^2-i\epsilon}\,
       d\sigma
  ~~\text{with}~~
   \rho_f(\sigma)=\frac{1}{\pi}\,
                   \textbf{Im}\,
                    \big[f(-\sigma)\big]\,.
 \end{eqnarray}
Then in the one-loop approximation for the running coupling its spectral density is
 $\rho_1(\sigma)=1/\sqrt{L_\sigma^2+\pi^2}$ and
\begin{subequations}
 \label{eq:A.U}
 \begin{eqnarray}
  \label{eq:A_1}
 \mathcal A_1[L]
  &=& \int_0^{\infty}\!\frac{\rho_1(\sigma)}{\sigma+Q^2}\,d\sigma\
   =\ \frac{1}{L} - \frac{1}{e^L-1}\,,~\\
 \label{eq:U_1}
 {\mathfrak A}_1[L_s]
  &=& \int_s^{\infty}\!\frac{\rho_1(\sigma)}{\sigma}\,d\sigma\
   =\ \frac{1}{\pi}\,\arccos\frac{L_s}{\sqrt{\pi^2+L_s^2}}\,,~
 \end{eqnarray}
\end{subequations}
whereas analytic images of the higher powers ($n\geq2, n\in\mathbb{N}$) are:
\begin{eqnarray}
 \label{eq:recurrence}
 {\mathcal A_n[L] \choose \mathfrak A_n[L_s]}
  &=& \frac{1}{(n-1)!}\left( -\frac{d}{d L}\right)^{n-1}
      {\mathcal A_{1}[L] \choose \mathfrak A_{1}[L_s]}\,.
\end{eqnarray}

In the standard QCD PT we have also:\\
(i) the factorization procedure in QCD
    that gives rise to the appearance of logarithmic factors of the type:
     $a_s^\nu[L]\,L$;~\footnote{%
     First indication that a special ``analytization'' procedure
is needed to handle these logarithmic terms appeared in~\cite{KS01}.}\\
(ii) the RG evolution
     that generates evolution factors of the type:
     $B(Q^2)=\left[Z(Q^2)/Z(\mu^2)\right]$ $B(\mu^2)$,
     which reduce in the one-loop approximation to\\
     $Z(Q^2) \sim a_s^\nu[L]$ with $\nu=\gamma_0/(2b_0)$
     being a fractional number.\\
All these means we need to construct analytic images of new functions:
$\displaystyle a_s^\nu,~a_s^\nu\,L^m, \ldots$\,.

In the one-loop approximation
using recursive relation (\ref{eq:recurrence})
we can obtain explicit expressions for
${\mathcal A}_{\nu}[L]$
and ${\mathfrak A}_{\nu}[L]$:
\begin{eqnarray}
 {\mathcal A}_{\nu}[L]
  = \frac{1}{L^\nu}
  - \frac{F(e^{-L},1-\nu)}{\Gamma(\nu)}\,;
 ~
 {\mathfrak A}_{\nu}[L]
  = \frac{\text{sin}\left[(\nu -1)\arccos\left(\frac{L}{\sqrt{\pi^2+L^2}}\right)\right]}
         {\pi(\nu -1) \left(\pi^2+L^2\right)^{(\nu-1)/2}}\,.~
\end{eqnarray}
Here $F(z,\nu)$ is the reduced Lerch transcendental function,
which is an analytic function in $\nu$.
They have very interesting properties,
which we discussed extensively in our previous papers~\cite{BMS-APT,AB08}.

Construction of FAPT with fixed number of quark flavors, $N_f$,
is a two-step procedure:
we start with the perturbative result $\left[a_s(Q^2)\right]^{\nu}$,
generate the spectral density $\rho_{\nu}(\sigma)$ using Eq.\ (\ref{eq:An.SD}),
and then obtain analytic couplings
${\mathcal A}_{\nu}[L]$ and ${\mathfrak A}_{\nu}[L]$ via Eqs.\ (\ref{eq:A.U}).
Here $N_f$ is fixed and factorized out.
We can proceed in the same manner for $N_f$-dependent quantities:
$\left[\alpha_s^{}(Q^2;N_f)\right]^{\nu}$
$\Rightarrow$
$\bar{\rho}_{\nu}(\sigma;N_f)=\bar{\rho}_{\nu}[L_\sigma;N_f]
 \equiv\rho_{\nu}(\sigma)/\beta_f^{\nu}$
$\Rightarrow$
$\bar{\mathcal A}_{\nu}^{}[L;N_f]$ and $\bar{\mathfrak A}_{\nu}^{}[L;N_f]$ ---
here $N_f$ is fixed, but not factorized out.\footnote{
Remind here that $\beta_f=b_0(N_f)/(4\pi)$.}

Global version of FAPT,
which takes into account heavy-quark thresholds,
is constructed along the same lines
but starting from global perturbative coupling
$\left[\alpha_s^{\,\text{\tiny glob}}(Q^2)\right]^{\nu}$,
being a continuous function of $Q^2$
due to choosing different values of QCD scales $\Lambda_f$,
corresponding to different values of $N_f$.
We illustrate here the case of only one heavy-quark threshold
at $s=m_4^2$,
corresponding to the transition $N_f=3\to N_f=4$.
Then we obtain the discontinuous spectral density
\begin{eqnarray}
 \label{eq:global_PT_Rho_4}
  \rho_n^\text{\tiny glob}(\sigma)
   = \theta\left(L_\sigma<L_{4}\right)\,
       \bar{\rho}_n\left[L_\sigma;3\right]
    + \theta\left(L_{4}\leq L_\sigma\right)\,
       \bar{\rho}_n\left[L_\sigma+\lambda_4;4\right]\,,~~~
\end{eqnarray}
with $L_{\sigma}\equiv\ln\left(\sigma/\Lambda_3^2\right)$,
$L_{f}\equiv\ln\left(m_f^2/\Lambda_3^2\right)$
and
$\lambda_f\equiv\ln\left(\Lambda_3^2/\Lambda_f^2\right)$ for $f=4$,
which is expressed in terms of fixed-flavor spectral densities
with 3 and 4 flavors,
$\bar{\rho}_n[L;3]$ and $\bar{\rho}_n[L+\lambda_4;4]$.
However it generates the continuous Minkowskian coupling
\begin{eqnarray}
 {\mathfrak A}_{\nu}^{\text{\tiny glob}}[L_s]
  \!&\!=\!&\!
    \theta\left(L_s\!<\!L_4\right)
     \Bigl(\bar{{\mathfrak A}}_{\nu}^{}[L_s;3]
          -\bar{{\mathfrak A}}_{\nu}^{}[L_4;3]
          +\bar{{\mathfrak A}}_{\nu}^{}[L_4+\lambda_4;4]
     \Bigr)
  \nonumber\\
  \!&\!+\!&\!
    \theta\left(L_4\!\leq\!L_s\right)\,
     \bar{{\mathfrak A}}_{\nu}^{}[L_s+\lambda_4;4]\,.
 \label{eq:An.U_nu.Glo.Expl}
\end{eqnarray}
and the analytic Euclidean coupling ${\cal A}_{\nu}^{\text{\tiny glob}}[L]$
(for more detail see in~\cite{AB08}).

\section{Resummation in the one-loop APT and FAPT}
\label{sec:Resum.FAPT}
We consider now the perturbative expansion
of a typical physical quantity,
like the Adler function and the ratio $R$,
in the one-loop APT.
Due to the limited space of our presentation
we provide all formulas only
for quantities in Minkowski region:
\begin{eqnarray}
 \label{eq:APT.Series}
  \mathcal R[L]
   = d_0
   + \sum_{n=1}^{\infty}
      d_n\,\mathfrak A_{n}[L]\,.
\end{eqnarray}
We suggest that there exist the generating function $P(t)$
for coefficients $\tilde{d}_n=d_n/d_1$:
\begin{equation}
 \tilde{d}_n
  =\int_{0}^\infty\!\!P(t)\,t^{n-1}dt
   ~~~\text{with}~~~
   \int_{0}^\infty\!\!P(t)\,d t = 1\,.
 \label{eq:generator}
\end{equation}
To shorten our formulae, we use for the integral
$\int_{0}^{\infty}\!\!f(t)P(t)dt$
the following notation:
$\langle\langle{f(t)}\rangle\rangle_{P(t)}$.
Then coefficients $d_n = d_1\,\langle\langle{t^{n-1}}\rangle\rangle_{P(t)}$
and as has been shown in~\cite{MS04}
we have the exact result for the sum in (\ref{eq:APT.Series})
\begin{eqnarray}
 \label{eq:APT.Sum.DR[L]}
  \mathcal R[L]
   = d_0 + d_1\,\langle\langle{\mathfrak A_1[L-t]}\rangle\rangle_{P(t)}\,.
\end{eqnarray}
The integral in variable $t$ here has a rigorous meaning,
ensured by the finiteness of the coupling  $\mathfrak A_1[t] \leq 1$
and fast fall-off of the generating function $P(t)$.

In our previous publications~\cite{AB08,BM08}
we have constructed generalizations of these results,
first, to the case of the global APT,
when heavy-quark thresholds are taken into account.
Then one starts with the series
of the type (\ref{eq:APT.Series}),
where $\mathfrak A_{n}[L]$
are substituted by their global analogs
$\mathfrak A_{n}^\text{\tiny glob}[L]$
(note that due to different normalizations of global
 couplings, $\mathfrak A_{n}^\text{\tiny glob}[L]\simeq\mathfrak A_{n}[L]/\beta_f$,
 the coefficients $d_n$ should be also changed).
Then
\begin{eqnarray}
 \mathcal R^\text{\tiny glob}[L]
  =  d_0
 \!\!&\!+\!&\!\! d_1 \langle\langle{
          \theta(L\!<\!L_4)
           \left[\Delta_{4}\bar{\mathfrak A}_{1}[t]
                +\bar{\mathfrak A}_{1}\!\Big[L\!-\!\frac{t}{\beta_3};3\Big]
           \right]     }\rangle\rangle_{P(t)}\nonumber\\
 \!\!&\!+\!&\!\! d_1 \langle\langle{
          \theta(L\!\geq\!L_4)
           \bar{\mathfrak A}_{1}\!\Big[L\!+\!\lambda_4-\!\frac{t}{\beta_4};4\Big]
                       }\rangle\rangle_{P(t)}\,;~~~
 \label{eq:sum.R.Glo.4}
\end{eqnarray}
where $\Delta_4\bar{\mathfrak A}_\nu[t]\equiv
  \bar{\mathfrak A}_\nu\!\Big[L_4+\lambda_{4}-t/\beta_4;4\Big]
 -\bar{\mathfrak A}_\nu\!\Big[L_3-t/\beta_3;3\Big]$.

The second generalization has been obtained for the case
of the global FAPT.
Then the starting point is the series of the type
$\sum_{n=0}^{\infty} d_n\,\mathfrak A_{n+\nu}^\text{\tiny glob}[L]$
and the result of summation is a complete analog of Eq.\ (\ref{eq:sum.R.Glo.4})
with substitutions
\begin{eqnarray}
 \label{eq:P_nu(t)}
  P(t)\Rightarrow P_{\nu}(t) =
   \int_0^{1}\!P\left(\frac{t}{1-x}\right)
    \frac{\nu\,x^{\nu-1}dx}
         {1-x}\,,
\end{eqnarray}
$d_0\Rightarrow d_0\,\bar{\mathfrak A}_{\nu}[L]$,
$\bar{\mathfrak A}_{1}[L-t]\Rightarrow
 \bar{\mathfrak A}_{1+\nu}[L-t]$,
and
$\Delta_4\bar{\mathfrak A}_{1}[t]\Rightarrow
 \Delta_4\bar{\mathfrak A}_{1+\nu}[t]$.
All needed formulas have been also obtained
in parallel for the Euclidean case.

\section{Applications to Higgs boson decay}
\label{sec:Appl.Higgs}
Here we analyze the Higgs boson decay to a $\bar{b}b$ pair.
For its width we have
\begin{eqnarray}
 \Gamma(\text{H} \to b\bar{b})
  = \frac{G_F}{4\sqrt{2}\pi}\,
     M_{H}\,
      \widetilde{R}_\text{\tiny S}(M_{H}^2)
  ~\text{with}~
  \widetilde{R}_\text{\tiny S}(M_{H}^2)
  \equiv m^2_{b}(M_{H}^2)\,R_\text{\tiny S}(M_{H}^2)
 \label{eq:Higgs.decay.rate}
\end{eqnarray}
and
$R_\text{\tiny S}(s)$
is the $R$-ratio for the scalar correlator,
see for details in~\cite{BMS-APT,BCK05}.
In the one-loop FAPT this generates the following
non-power expansion\footnote{%
Appearance of denominators $\pi^n$ in association
with the coefficients $\tilde{d}_n$
is due to $d_n$ normalization.}:
\begin{eqnarray}
 \widetilde{\mathcal R}_\text{\tiny S}[L]
   =  3\,\hat{m}_{(1)}^2\,
      \Bigg\{\mathfrak A_{\nu_{0}}^{\text{\tiny glob}}[L]
          + d_1^\text{\,\tiny S}\,\sum_{n\geq1}
             \frac{\tilde{d}_{n}^\text{\,\tiny S}}{\pi^{n}}\,
              \mathfrak A_{n+\nu_{0}}^{\text{\tiny glob}}[L]
      \Bigg\}\,,
 \label{eq:R_S-MFAPT}
\end{eqnarray}
where $\hat{m}_{(1)}^2=8.45$~GeV$^2$ is the RG-invariant
of the one-loop $m^2_{b}(\mu^2)$ evolution
$m_{b}^2(Q^2) = \hat{m}_{(1)}^2\,\alpha_{s}^{\nu_{0}}(Q^2)$
with $\nu_{0}=2\gamma_0/b_0(5)=1.04$ and
$\gamma_0$ is the quark-mass anomalous dimension
(for a discussion --- see in~\cite{KK08}).

\begin{figure}[ht]
 \centerline{\includegraphics[width=0.48\textwidth]{
             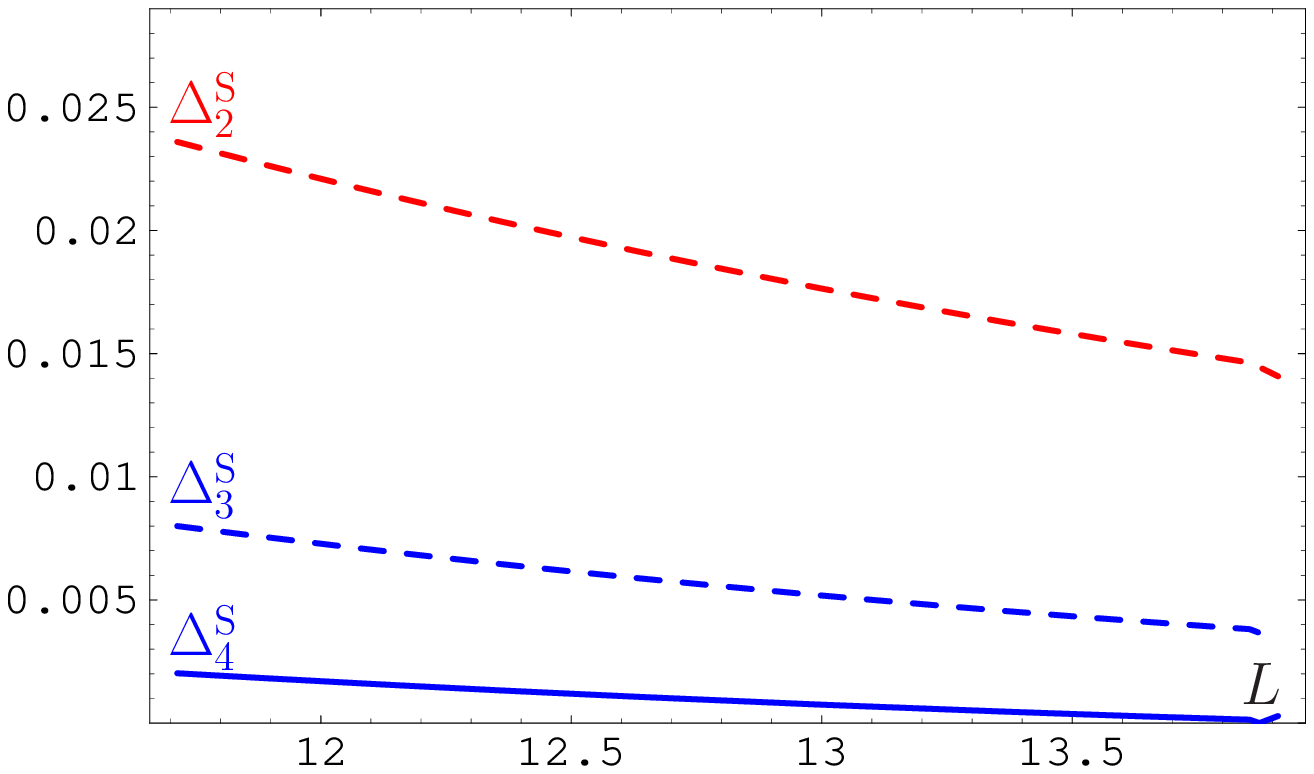}
          ~~~\includegraphics[width=0.47\textwidth]{
             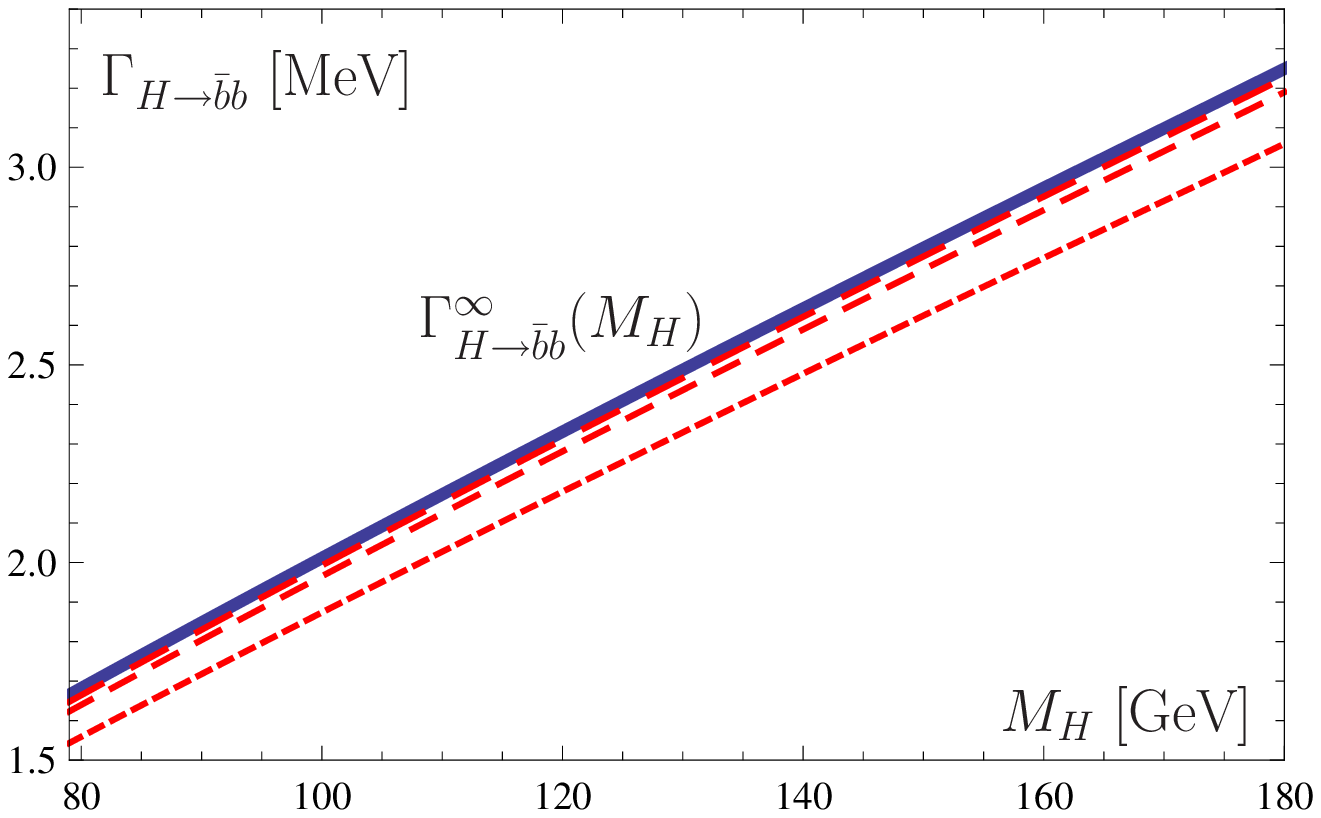}
  \vspace*{-1mm}}
   \caption{Left: The relative errors  $\Delta^\text{S}_N[L]$, $N=2, 3$
    and $4$, of the truncated FAPT in comparison with the exact summation result,
    Eq.\ (\ref{eq:R_S.Sum}).
    Right: The width $\Gamma_{H\to b\bar{b}}^{\infty}$ as a function of the  Higgs
    boson mass $M_{H}$ in the resummed FAPT (solid line).
   \label{fig:Higgs}}
\end{figure}
We take for the generating function $P(t)$
the Lipatov-like model of~\cite{BM08}
with $\left\{c=2.4,~\beta=-0.52\right\}$
\begin{eqnarray}
\label{eq:Higgs.Model}
  \tilde{d}_{n}^\text{\,\tiny S}
   = c^{n-1}\frac{\Gamma (n+1)+\beta\,\Gamma (n)}{1+\beta}\,;\quad
  P_\text{\tiny S}(t)
  = \frac{(t/c)+\beta}{c\,(1+\beta)}\,e^{-{t/c}}\,.
\end{eqnarray}
It gives a very good prediction for
$\tilde{d}_{n}^\text{\,\tiny S}$ with $n=2, 3, 4$,
calculated in the QCD PT~\cite{BCK05}:
$7.50$, $61.1$, and  $625$
in comparison with
$7.42$, $62.3$, and  $620$.
Then we apply FAPT resummation technique
to estimate
how good is FAPT
in approximating the whole sum $\widetilde{\mathcal R}_\text{\tiny S}[L]$
in the range $L\in[11.5,13.7]$
which corresponds to the range
$M_H\in[60,180]$~GeV$^2$
with $\Lambda^{N_f=3}_{\text{QCD}}=189$~MeV
and ${\mathfrak A}^{\text{\tiny glob}}_{1}(m_Z^2)=0.122$.
In this range we have ($L_6=\ln(m_t^2/\Lambda_3^2)$)
\begin{eqnarray}
 \frac{\widetilde{\mathcal R}_\text{\tiny S}[L]}
      {3\,\hat{m}_{(1)}^2}
  = {\mathfrak A}^\text{\tiny glob}_{\nu_{0}}[L]
   + \frac{d_{1}^\text{\,\tiny S}}{\pi}\,
      \langle\langle{\bar{\mathfrak A}_{1+\nu_{0}}\!
                      \Big[L\!+\!\lambda_5\!-\!\frac{t}{\pi\beta_5};5\Big]
               \!+\! \Delta_{6}\bar{\mathfrak A}_{1+\nu_{0}}
                      \left[\frac{t}{\pi}\right]
                      }\rangle\rangle_{P_{\nu_{0}}^\text{\,\tiny S}}~
 \label{eq:R_S.Sum}
\end{eqnarray}
with $P_{\nu_{0}}^\text{\,\tiny S}(t)$ defined via Eqs.\ (\ref{eq:Higgs.Model})
and (\ref{eq:P_nu(t)}).
Now we analyze the accuracy of the truncated FAPT expressions
\begin{eqnarray}
 \label{eq:FAPT.trunc}
 \widetilde{\mathcal R}_\text{\tiny S}[L;N]
  &=& 3\,\hat{m}_{(1)}^2\,
       \left[{\mathfrak A}_{\nu_{0}}^{\text{\tiny glob}}[L]
           + d_1^\text{\,\tiny S}\,\sum_{n=1}^{N}
              \frac{\tilde{d}_{n}^\text{\,\tiny S}}{\pi^{n}}\,
               {\mathfrak A}_{n+\nu_{0}}^{\text{\tiny glob}}[L]
       \right]
\end{eqnarray}
and compare them with the total sum
$\widetilde{\mathcal R}_\text{\tiny S}[L]$
in Eq.\ (\ref{eq:R_S.Sum})
using relative errors
$\Delta_N^\text{S}[L]=1-\widetilde{\mathcal R}_\text{\tiny S}[L;N]/\widetilde{\mathcal R}_\text{\tiny S}[L]$.
In the left panel of Fig.\ \ref{fig:Higgs}
we show these errors for $N=2$, $N=3$, and $N=4$
in the analyzed range of $L\in[11,13.8]$.
We see that already $\widetilde{\mathcal R}_\text{\tiny S}[L;2]$
gives accuracy of the order of 2.5\%,
whereas $\widetilde{\mathcal R}_\text{\tiny S}[L;3]$
of the order of 1\%.
That means that there is no need to calculate further corrections:
at the level of accuracy of 1\% it is quite enough to take into account
only coefficients up to $d_3$.
This conclusion is stable
with respect to the variation of parameters
of the model $P_\text{\tiny S}(t)$
and is in a complete agreement
with Kataev--Kim conclusion~\cite{KK08}.

\section{Conclusions}
\label{sec:Concl}
In this report we described the resummation approach
in the global versions of the one-loop APT and FAPT
and argued that it produces finite answers,
provided the generating function $P(t)$
of perturbative coefficients $d_n$ is known.
The main conclusion is:
To achieve an accuracy of the order of 1\%
we do not need to calculate more than four loops and
$d_4$ coefficients are needed only to estimate
corresponding generating functions $P(t)$.

\section*{Acknowledgements}
This work was supported in part by
the Russian Foundation for Fundamental Research,
grants No.\ ü~07-02-91557 and 08-01-00686,
the BRFBR--JINR Cooperation Programme,
contract No.\ F08D-001, and
the Heisenberg--Landau Programme under grant 2009.


\end{document}